\documentclass[12pt]{article}

\usepackage{graphicx}   \usepackage{jtb}
\begin{document}

\title{The Dynamics of Acute Inflammation} \author{Rukmini
Kumar$^{1}$, Gilles Clermont$^2$, Yoram Vodovotz$^{3}$ and Carson
C. Chow$^{4}$\footnote{Corresponding  
Author, Present Address: Laboratory for Biological Modeling, NIDDK,
NIH, Bethesda, MD 20892}\\
$^1$Departments of Physics and Astronomy \\
$^2$Department of Critical Care Medicine\\
$^3$Department of Surgery\\
$^4$ Department of Mathematics \\ 
University of Pittsburgh, Pittsburgh, PA 15260}
%\date{Type Date Here}

\maketitle
%
%\documentclass[12]{report}
%\usepackage{amsmath}
%\usepackage{latexsym}
%\usepackage{psfig}
%\usepackage{graphics}
%\usepackage[dvips]{graphicx}
%\hbadness=10000 \tolerance=1600
%\setlength{\textwidth}{6.5 in}
%\setlength{\oddsidemargin}{0.0 in}
%\setlength{\evensidemargin}{0.5 in}
%\setlength{\topmargin}{-0.5625in}
%\setlength{\textheight}{9.0 in}
%\setlength{\headheight}{0.25in}
%\setlength{\headsep}{0.4in}
%\renewcommand{\baselinestretch}{1.5}
%\pagestyle{plain} 

\abstract{When the body is infected, it mounts an
\emph{acute inflammatory response} to rid itself of the pathogens and
restore health. Uncontrolled acute inflammation due to infection is
defined clinically as \emph{sepsis} and can culminate in organ
failure and death. We consider a three dimensional ordinary
differential equation 
model of inflammation consisting of a pathogen, and two inflammatory
mediators. The model reproduces the healthy outcome and diverse
negative outcomes, depending on initial conditions and parameters.  
We analyze the various bifurcations between the different outcomes
when key parameters are changed and suggest various therapeutic
strategies.  We suggest 
that the clinical condition of sepsis can arise from several distinct
physiological states, each of which requires a different treatment
approach.}  

\newpage
\section{Introduction}
  The initial  response of  the body  to an  infection or  trauma is
called   the   acute  inflammatory   response.   This  response   is
non-specific and  is the first line  of defense of  the body against
danger~\cite{janeway}.  It  consists  of  a  coordinated  local  and
systemic   mobilization  of   immune,  endocrine   and  neurological
mediators. In a healthy  response, the inflammatory response becomes
activated, clears the pathogen (in the event of infection), begins a
repair process  and abates. However, inflammation  itself can damage
otherwise   healthy  cells  which   could  then   further  stimulate
inflammation.  This  runaway inflammation can lead  to organ failure
and death. Systemic inflammation  accompanied by infection, based on
its      clinical       manifestations,      is      defined      as
sepsis~\cite{sepsisdefnold}.   Sepsis   is   a   common   and 
frequently  fatal  condition, with  750,000  cases  annually in  the
United States alone in  1995~\cite{Angus}.

Though much  has been learned about the molecular and physiological
pathways of the acute  inflammatory response, this knowledge has not
led  to  many  effective  therapies  against  sepsis.  The sole
approved drug therapy for severe sepsis is
activated  Protein-C, which only  reduced  mortality by  6\% compared  with
controls in clinical trials~\cite{APC,review}. 
One  reason  for  the lack  of  effective
treatments may be that the  complex nature of the inflammatory response
renders the  effect of  targeting isolated components  of inflammation
difficult  to  predict.   Thus,  mathematical modeling  may  provide
insights into  the global dynamics of the  inflammatory process from
which therapies may be developed.  We propose that simple models of the
acute  inflammatory  response can exhibit various outcomes  and
facilitate an understanding of  the complex interactions between the
various components of the response.

We   present  a   simple  3-dimensional  model   of  the
inflammatory  response  to  infection  that captures  the  following
clinically  relevant  scenarios:  a  \emph{healthy} response  where
pathogen   is  cleared   and  the   body  returns   to  homeostasis,
\emph{recurrent infection} where  inflammation is inadequate and the
pathogen   cannot   be   completely   eliminated,   \emph{persistent
infectious inflammation} where  the pathogen levels and inflammation
are   high,  \emph{persistent  non-infectious   inflammation}  where
pathogen  is  cleared  but  inflammation persists  and  \emph{severe
immuno-deficiency} where  pathogen has  grown to saturation  but the
inflammatory response  is very low.  The model suggests that sepsis
is a multifaceted disease and  narrowly targeted interventions are
unlikely to succeed.
 
\section{Reduced Model of Acute Inflammation}

Invading pathogens such as bacteria are rapidly detected by the body and
an acute inflammatory response ensues.
Among the first responders are
phagocytic immune cells such as neutrophils and macrophages.  These
immune cells detect the bacterial cell components, become ``activated"
and begin to release pro-inflammatory cytokines such as Tumor Necrosis
Factor-$\alpha$ (TNF-$\alpha$), 
Interleukins (IL)-1, IL-6, IL-8  and High Motility Group Box-1
(HMGB-1) that activate more immune cells and recruit them to the 
sites of the infection~\cite{review}. In addition, anti-inflammatory
mediators such 
as IL-10 and Transforming Growth Factor-$\beta$
(TGF-$\beta$) are also released which inhibit the 
production of the pro-inflammatory mediators.  The activated 
phagocytic cells kill bacteria directly by engulfment and secretion of
toxic chemicals such as oxygen free radicals~\cite{janeway}.  These
substances can damage otherwise healthy cells.  These damaged or
dysfunctional cells can then induce more inflammation~\cite{decide1}.
 
Ideally, the inflammatory response eliminates the pathogen and then
subsides. In some cases, the response might not be strong enough to
clear the pathogen. In other cases, a positive feedback loop could
arise between the early and late pro-inflammatory waves leading to a
non-abating inflammatory response.  Clinically, a sustained acute
inflammatory response is manifested as septic shock and could
culminate in organ failure and death.  Though this is a simplification
of the pathogenesis of sepsis, the idea is supported by the
persistence of high levels of pro-inflammatory cytokines in
non-survivors of sepsis~\cite{pinsky,abraham}.

A simplified picture of the acute inflammatory response is that
an infectious pathogen triggers early pro-inflammatory responders
which attempt to kill the pathogen. The early inflammatory mediators
then activate later inflammatory mediators which can further excite
the early mediators.  This is the basis of our model which consists of
three variables: 1) a pathogen $p$, which is an instigator of the
innate immune response; 2) an early pro-inflammatory mediator $m$,
which can be thought of as representing the combined effects of immune
cells such as macrophages and neutrophils together with early
pro-inflammatory mediators such as TNF-$\alpha$ and IL-1; 3) a late
pro-inflammatory mediator $l$ which represents a late pro-inflammatory
feedback.  This is a combined effect of cytokines such as IL-6, HMGB-1
and stimulatory effects of tissue damage and dysfunction.  Although
our model is extremely simple, it captures qualitatively all the
salient features of a more complicated and biologically faithful model
that is currently being developed~\cite{ccm}.

We consider mass-action type kinetics in a well-mixed volume.  The
dynamics obey:
\begin{eqnarray}
\frac{dp}{dt}&=&k_pp(1-p)-k_{pm}mp\label{model1}\\
\frac{dm}{dt}&=&(k_{mp}p+l)m(1-m)-m\label{model2}\\
\frac{dl}{dt}&=&k_{lm}f(m)-k_ll \label{model3}
\end{eqnarray}
where
\begin{equation}
f(m)=1+\tanh \left(\frac{m-\theta}{w}\right),
\label{coupling}
\end{equation}
$\theta$ is an activation threshold and $w$ is an activation width.
All the variables and parameters are non-negative.  The pathogen $p$
obeys logistic growth and is killed when it interacts with $m$.  The
presence of $p$ or $l$ and $m$ will stimulate the growth of $m$ which
also has an intrinsic death rate.  This growth saturates as $m$
increases towards unity mimicking the effects of cell depletion and
anti-inflammatory cytokines.  The late mediator $l$ is recruited by
$m$ through a sigmoidal coupling function (\ref{coupling}) and is
cleared with the rate of $k_l$. In summary, the dynamics
obey predator-prey  dynamics with a delayed response.

We note that the specific nature of the interactions of our model are
not essential for the  qualitative dynamics we find.  As an example,
the factor $1-m$ in (\ref{model2}) may be replaced by $1/(1+m)$ or the
form of the coupling function (\ref{coupling}) can be changed without affecting
our conclusions.  The important element is that there be an early and
late inflammatory mediator with saturating kinetics.

\section{Dynamics of the Model}

\subsection{Numerical examples}

The model (\ref{model1}) - (\ref{model3}) exhibits behavior reminiscent
of what is observed in clinical settings.  Given an initial condition
of $p>0$, $m>0$ and $l>0$, $p$ grows, inducing $m$ and $l$ to grow and
attempt  to eliminate $p$.  Depending on initial conditions and
parameters, the ensuing orbits either approach stable fixed points or
undergo oscillations, each having a physiological interpretation.  We
note from (\ref{model2}) that a non-zero positive initial value for
$m$ is necessary to generate an inflammatory response. The background
level of late mediators is given by $l^{0}=(k_{lm}/k_l)f(0)$ and is
non-zero for our choice of parameters.  Both these properties are
consequences of our particular choice for the form of the model and
not essential for the qualitative results we find.

We must adopt one crucial element before we can interpret our results.
In a healthy response to infection, the inflammatory response should
become activated, eliminate the bacteria and return to rest.  However,
in model (\ref{model1})-(\ref{model3}), while $p$ can diminish
exponentially fast, it can never become completely eliminated in a 
finite time.  Thus a consequence is that if $m$ returns to rest after
an inflammatory episode, $p$ can re-grow from an infinitesimal
quantity.  This is an artifact of the model that arises because the
approximation of $p$ as a continuous variable breaks down when the
population becomes small and the discreteness of the pathogen number
becomes important.  In this regime, a stochastic or agent based model
where the pathogen can be completely eliminated is more appropriate.
We finesse the discreteness problem by introducing a threshold for
pathogen level.  When $p$ falls below this threshold, we consider it
to be completely cleared.   We propose that there is an effective threshold
representing a single pathogen particle below which,  on average, the
pathogen population is eliminated.  We will show how to calculate this
threshold explicitly for various models in a future publication.

We consider numerical examples for various values of $k_{pm}$,
$k_{mp}$, and $k_{lm}$.  In Section \ref{fpandb}, we show that these are 
natural bifurcation parameters of the system.  
The other parameters are fixed at $k_{pg}=3$,
$k_{l}=1$, $\theta=1$ and $w=0.5$.   
We show the effects of varying 
$\theta$ and $w$ in Appendix \ref{couplingcurve}.  Numerical
simulations and bifurcation plots were generated with
XPPAUT~\cite{bardxpp}.

A {\em healthy response} to infection as seen in
Fig.~\ref{fig:healthy} corresponds to an orbit that spirals outwards
so that $p$ falls below threshold during the oscillation.  The
pathogen is then completely cleared which allows the inflammatory
response to relax back to rest. 

\begin{figure} 
\begin{center}
\includegraphics[angle=270,scale=0.4]{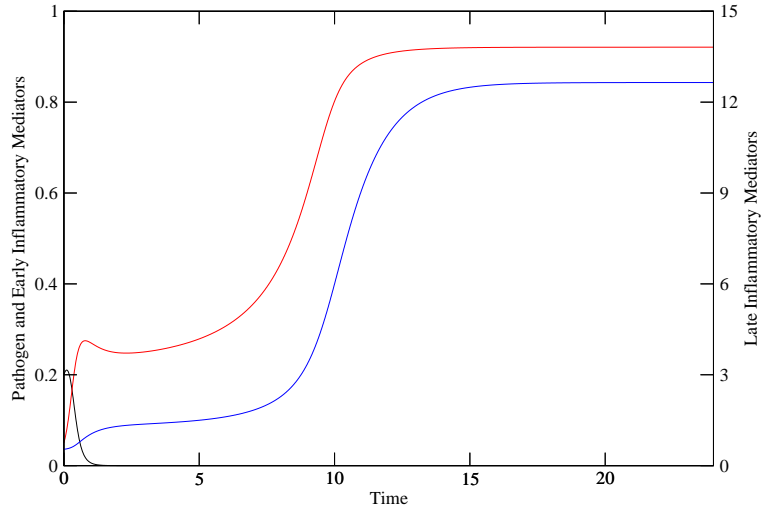}
\end{center}
\caption{In {\it persistent non-infectious inflammation},  $p$ (black) is
eliminated but $m$ (red) and $l$ (blue) remain elevated.
Parameters used are $k_{pm}=30, k_p=3, k_{mp}=25, k_{lm}=15,
k_l=1$. Initial conditions are $ p(0)=0.2, m(0)=0.05, l(0)=0.539$.}
\label{fig:shock}
\end{figure}

In Fig.~\ref{fig:shock}, the same
parameters are used but the initial pathogen load is higher so that
instead of returning to background levels, the inflammatory mediators
are over-excited and remain elevated.  We relate this situation to a
state of {\em persistent non-infectious inflammation} where even
though the pathogen is cleared, the inflammatory response does not
abate. In our model, this state is a fixed point but in a real
organism we expect that if this condition continued, it would
eventually lead to multiple organ failure and
death~\cite{noinfsepsis}.  

\begin{figure} 
\begin{center}
\includegraphics[angle=270,scale=0.4]{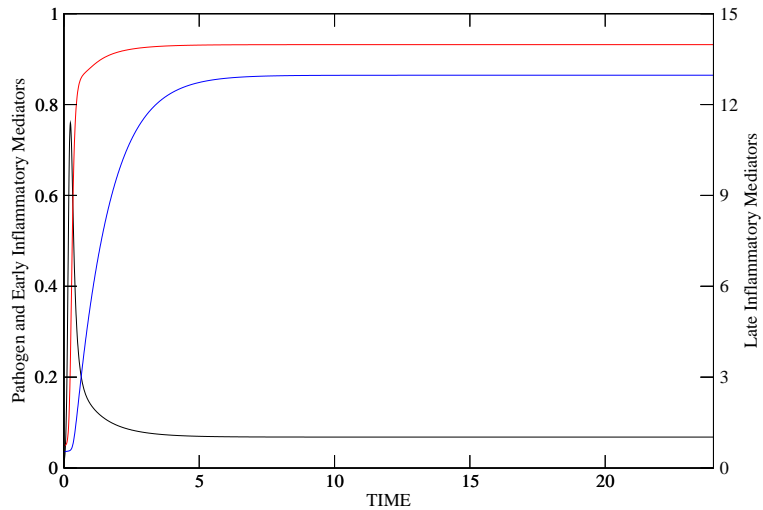}
\end{center}
\caption{In {\it persistent infectious inflammation}, $p$ (black) cannot be
eliminated and $m$ (red) and $l$ (blue) remain elevated (as in
severe sepsis). Parameters used are
$k_{pm}=3, k_p=3, k_{mp}=25, k_{lm}=15, k_l=1$. Initial conditions are
$p(0)=0.01, m(0)=0.05, l(0)=0.539$.}
\label{fig:sepsis}
\end{figure}

Starting again from a healthy situation, if the pathogen
susceptibility to the host's defenses ($k_pm$)  is decreased, we can  
enter a domain of {\em persistent infectious 
inflammation} where the inflammatory response is high but the pathogen
still cannot be cleared as seen in Fig.~\ref{fig:sepsis}.  We would
equate this condition with a severe septic state where both infection
and inflammation are uncontrolled.  In this case the damage caused by
both the pathogen and inflammation are disrupting body function and if
unabated death will result. Patients with systemic inflammation, with
and without documented infection, 
are observed in clinical settings~\cite{epidem02}.

\begin{figure} 
\begin{center}
\includegraphics[angle=270,scale=0.4]{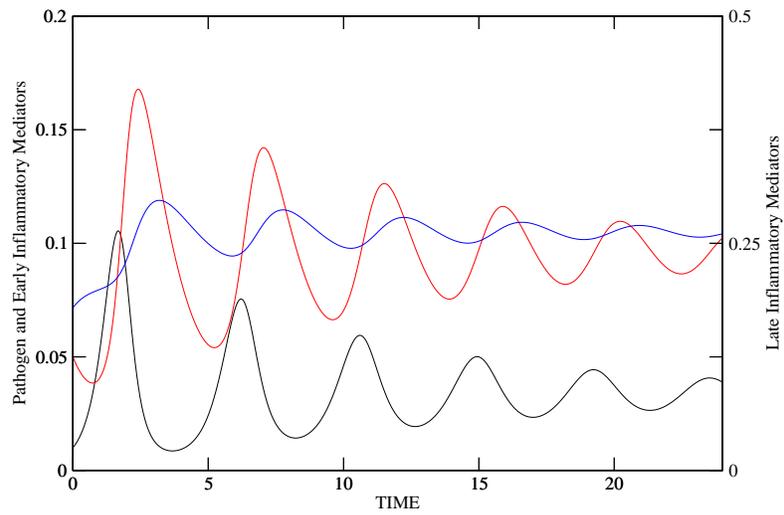}
\end{center}
\caption{In {\it recurrent infection}, $p$ (black) is low and $m$ (red) and $l$ (blue) remain above background values but not very
high (as in a low-grade infection). Parameters used are
$k_{pm}=30, k_p=3, k_{mp}=25, k_{lm}=5, k_l=1$. Initial conditions are
$p(0)=0.01, m(0)=0.05, l(0)=0.179$.}
\label{fig:recurr}
\end{figure}

If we reduce the recruitment rate of $l$ ($k_{lm}$), the healthy response
can be turned into one of recurrent infection as seen in
Fig.~\ref{fig:recurr}. (Changes in other parameters could also lead to
recurrent infection as we see in the next section).  In this case, low
levels of infection persist indefinitely.  This could be likened to
infection with tuberculosis, yeast infections or low-grade bacterial
infections that persist for long 
periods of time~\cite{lowgrade}.  Although, an organism could survive
this state for a 
long duration it may eventually succumb.

\begin{figure} 
\begin{center}
\includegraphics[angle=270,scale=0.4]{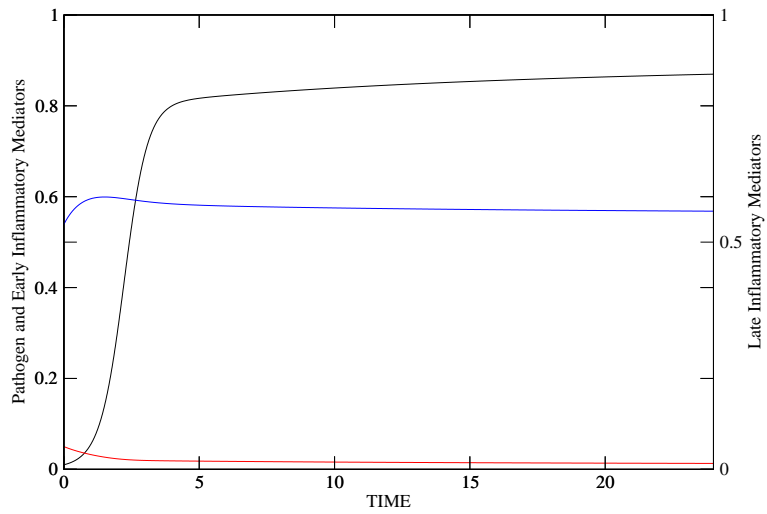}
\end{center}
\caption{In {\it severe immuno-deficiency},  $p$ (black) grows to saturation
and $m$ (red) and $l$ (blue) remains low or absent.  Parameters are
$k_{pm}=30, k_p=3, k_{mp}=0.4, k_{lm}=15, k_l=1, p(0)=0.01, m(0)=0.05, l(0)=0.539$.}
\label{fig:suppress}
\end{figure}

Finally, in Fig.~\ref{fig:suppress}, the activation rate of $m$ due to
$p$ ($k_{mp}$) is very  weak, and we have a state  of {\em
  immuno-suppression} or 
{\em  immuno-deficiency} where  the pathogen  grows to  saturation and
does not elicit any response from  the body.  This could happen if the
immune-system had  already been  compromised by previous  infection or
trauma and  then the organism is  reinfected.  Opportunistic bacterial
and fungal infections have  been observed in immuno-suppressed patient
populations such as HIV infected  patients, the elderly and those with
organ transplants~\cite{nosoc}.
These five scenarios are the only possible outcomes in the model.

\subsection{Fixed Points and Bifurcations}
\label{fpandb}

These various regimes and transitions are best understood by examining
the fixed points and associated bifurcations of the model
(\ref{model1})-(\ref{model3}).  The fixed points satisfy the following
conditions:
\begin{eqnarray}
0&=& p\left[p-1+\left(\frac{k_{pm}}{k_p} \right) m\right]
\label{cond1}\\
0&=&m\left[m-1+\left(\frac{1}{k_{mp}p+l}\right)\right]\label{cond2}\\
l&=&\frac{k_{lm}}{k_l}f(m)\label{cond3}
\end{eqnarray}
Substituting (\ref{cond3}) into (\ref{cond2}) and rearranging gives
\begin{equation}
0=m\left[p-\frac{1}{k_{mp}}\left(\frac{1}{1-m}-\frac{k_{lm}}{k_l}f(m)\right)\right].
\label{cond4}
\end{equation}
We consider three natural parameters $a\equiv k_{pm}/k_p$, $k_{mp}$,
and $b\equiv k_{lm}/k_l$, that appear in the fixed point conditions
(\ref{cond1}) and (\ref{cond4}). These three parameter combinations
represent the pathogen susceptibility to $m$ compared to its growth
rate (i.e. inverse of the pathogen virulence), the activation rate
of early responders $m$ due to $p$, and the effective recruitment rate of $l$
due to $m$, respectively.

\begin{figure} 
\begin{center}
\includegraphics[angle=270,scale=0.4]{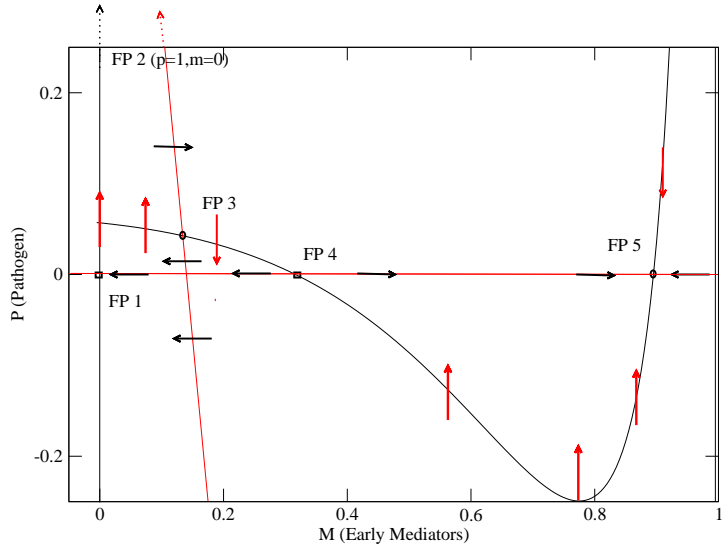}
\end{center}
\caption{Projection of fixed-point conditions onto the $p$-$m$ plane.
Projected directions of field-flow are also
marked. The red curves correspond to conditions $p=0$ and (\ref{fp2})
and the black curves correspond to conditions $m=0$ and 
(\ref{fp6}). The fixed points are at the intersections of black and
red curves  and the direction of flow along a curve reverses at  
fixed points. The projection of the phase-space for a typical set of
parameters is shown.  FP~1 and FP~4 are saddles. FP~3 is a spiral
which is stable or unstable based on the actual values of the
parameters. FP~5 is a stable fixed point. FP~2 (not shown) is also a
saddle in this case.}
\label{fig:nullclines}
\end{figure}

The intersections of conditions (\ref{cond1}) and (\ref{cond4}) give
the fixed points of the system.  The dependence of the fixed points
with the parameters is best observed as a projection in the $p-m$
plane as shown in   
Fig.~\ref{fig:nullclines} where the fixed point conditions
(\ref{cond1}) and (\ref{cond4}) are plotted. Lines $p=0$ and $m=0$ are
unaffected by changes in parameters. The line
\begin{equation}
p=1-a m \label{fp2}
\end{equation} 
is affected only by $a$ and sweeps across the $p-m$ plane as $a$ is decreased.
The curve
\begin{equation}
p=\frac{1}{k_{mp}}\left(\frac{1}{1-m}-b f(m)\right).
\label{fp6}
\end{equation} 
is affected only by the immune parameters $b$ and $k_{mp}$. It drops
below $p=0$ as $b$ is
increased. Changing $k_{mp}$ affects the height and angle at which
(\ref{fp6}) intersects (\ref{fp2}).

There are five fixed points which we have labeled from FP~1 to FP~5. 
FP~1 is given by $p=0$, $m=0$, and $l=(k_{lm}/k_l)f(0)$.  This
fixed point is always unstable because the pathogen is growing.  FP~2
is given by $p=1$, $m=0$, $l=(k_{lm}/k_l)f(0)$ and may be interpreted
as a severely \emph{immuno-deficient} state.  Here the pathogen has
grown to saturation, but there is no early immune response and the
late response remains at the background level.  This point is stable when the
early immune response is very weak as shown in Appendix~\ref{appfp2}.

\begin{figure} 
\begin{center}
\includegraphics[angle=270,scale=0.4]{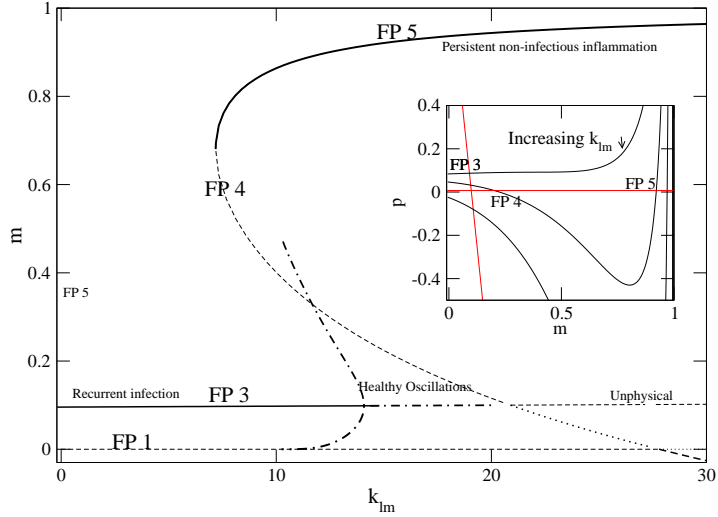}
\end{center}
\caption{Bifurcation plot showing outcome as $k_{lm}$  or $b=k_{lm}/k_l$
is increased. For low $k_{lm}$, FP 3 is a stable
spiral representing {\em recurrent infection}. As $k_{lm}$ is increased, FP
4 and 5 are created through a saddle-node bifurcation at $k_{lm} \approx
7$ and the {\em persistent non-infectious inflammation} is now a
possible outcome. At  $k_{lm}\approx 14$, FP 3 undergoes a
subcritical Hopf bifurcation and is surrounded
by an unstable limit cycle. The oscillations around the unstable limit
cycle are interpreted as {\em healthy} behavior and trajectories
inside the limit cycle spiral into FP 3 and are interpreted as {\em
recurrent infection}. {\it Healthy} behavior is lost  at $k_{lm} \approx 10$
due to a homoclinic bifurcation. Beyond the Hopf bifurcation, {\em
healthy} oscillations are still supported by FP 3 which is now an
unstable spiral. At  $k_{lm} \approx 21$, FP 3 meets FP 4 and
becomes unphysical through a transcritical bifurcation (as its $p<0$). Beyond this point, FP 5 is the only stable attractor and {\em persistent
non-infectious inflammation} is the only possible outcome of the
system.  (Inset :  In $p$-$m$ plane,
 (\ref{fp6}) descends as $k_{lm}$
is increased causing the above bifurcations.)}
\label{fig:varylt}
\end{figure}

FP 4 and FP 5 arise from a saddle node bifurcation when $p=0$ in
(\ref{fp6}). This can be achieved by increasing $b$ from zero as seen
in the bifurcation plot in Fig.~\ref{fig:varylt}.  FP~4 is always
unstable and never represents any physiologically relevant
scenario. When stable, FP~5 represents the {\em persistent
non-infectious inflammation} fixed point. FP~3 is given by the
intersection of the line (\ref{fp2}) with the curve (\ref{fp6}). FP~3
could represent {\em healthy}, {\em recurrent infection} or {\em
persistent infectious inflammation} states, depending on parameter
values. Below, we vary the three parameter combinations $a$, $k_{mp}$ 
and $b$ and examine the bifurcation plots.

\begin{figure} 
\begin{center}
\includegraphics[angle=270,scale=0.4]{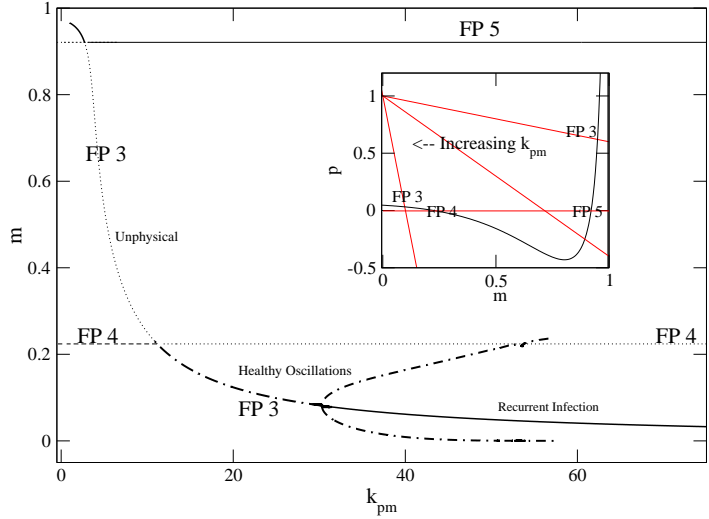}
\end{center}
\caption{Bifurcation plot showing outcome as $k_{pm}$ and hence
  $a=k_{pm}/k_p$ is increased. For low $k_{pm}$, FP 3 interpreted as
  {\em persistent infectious 
  inflammation} is the only fixed 
point of the system. As $k_{pm}$ is increased to
$\approx 2.7$, FP 5 ({\em persistent infectious inflammation})
becomes the attractor of the system. FP 3 becomes unphysical and
moves leftward in the $p$-$m$ plane (see inset). At $k_{pm}
\approx 11$, FP 3 undergoes a transcritical bifurcation with FP 4,
becoming physical again. It is an unstable spiral, oscillations
around which are interpreted as {\em healthy}.  On further increasing
$k_{pm}$, FP 3 undergoes a subcritical Hopf
bifurcation and is surrounded by an unstable limit-cycle, oscillations
around which are interpreted as {\em healthy}. Trajectories within the
limit-cycle, which spiral into FP 3, are interpreted as {\em recurrent
infection}. {\it Healthy} behavior is lost to a homoclinic bifurcation at
$k_{pm} \approx 57$. For large $k_{pm}$, FP 3 and FP 5 are
bistable (i.e.~{\em recurrent infection} and {\em persistent
non-infectious inflammation} are possible outcomes of the
system). (Inset : In the $p$-$m$ plane, line (\ref{fp2}) sweeps across
  the plane causing the above bifurcations.) } 
\label{fig:varyp}
\end{figure}

In the healthy scenario, $a$ is large enough so that FP~3 is an
unstable spiral and FP 5 is the only stable node ($k_{pm}=20$ in
Fig.~\ref{fig:varyp}). In this case, initial conditions close to FP 3
undergo oscillations that take the pathogen below the elimination
threshold and are interpreted as healthy trajectories.  A higher
pathogen initial condition for the same set of parameters takes the
trajectories directly to FP 5 and this behavior is interpreted as
persistent non-infectious inflammation. When $a$ is decreased,  the
slope of line (\ref{fp2}) decreases and it sweeps through FP 4
rendering it unstable through a transcritical bifurcation.  However,
$p$ is negative and thus FP 3 becomes unphysical ($k_{pm}=10$ in
Fig.~\ref{fig:varyp}). Hence FP 5 is the only stable fixed point for
all initial conditions in this regime.  As $a$ is decreased further FP
3 crosses FP 5 and they exchange stability through another
transcritical bifurcation.  FP 3 then becomes the only stable fixed
point and is the global attractor for all initial conditions.

As expected, starting from a healthy scenario,  the outcome of the
system becomes more and more severe with decreasing pathogen
susceptibility (i.e. increasing effective pathogen virulence). If we
were to increase 
susceptibility $a$ starting from the healthy scenario, FP 3 changes
into a stable spiral 
surrounded by an unstable limit cycle through a subcritical Hopf
bifurcation. Now, trajectories within the unstable limit cycle spiral
into the stable fixed point FP 3 resulting in {\em recurrent
infection} and unstable spiral trajectories outside the unstable limit
cycle will eventually dip below threshold and be interpreted as
healthy. As FP 5 is still the attractor when the initial $p$ or $m$ is
very high, this set of parameters could lead to three different
outcomes based on the initial conditions ($k_{pm}=40$ in
Fig.~\ref{fig:varyp}). Increasing $a$ further, increases the radius of
the unstable limit cycle until it undergoes a homoclinic bifurcation
when it collides with FP 1. Here FP 3 is a stable spiral and only
recurrent infection is supported. The model therefore predicts that
pathogen clearance is not possible for a range of high pathogen
susceptibility. However, as we increase susceptibility much further, FP
3 gets closer to the origin so that oscillations around it take $p$
below threshold and may be interpreted as healthy.

\begin{figure} 
\begin{center}
\includegraphics[angle=270,scale=0.4]{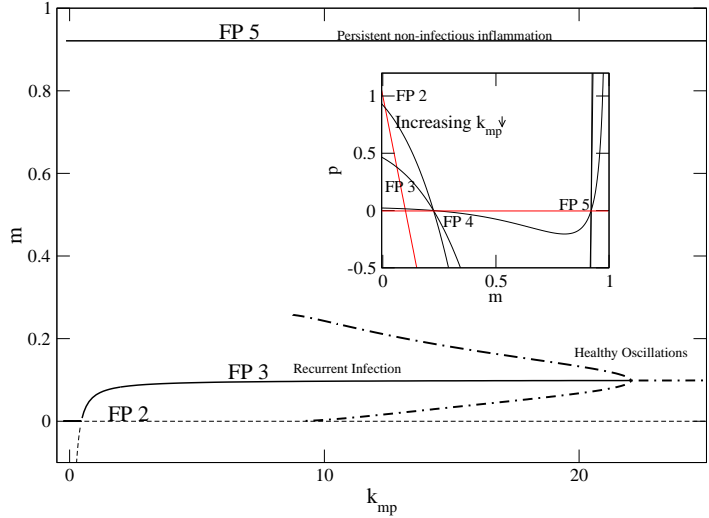}
\end{center}
\caption{Bifurcation plot showing outcome as $k_{mp}$ is
increased. For very low $k_{mp}(<0.5)$, FP~2 is 
stable and is interpreted as the {\em severely immuno-deficient}
state.  As $k_{mp}$ is increased, FP~3 becomes stable through a
transcritical bifurcation and {\em recurrent infection} is a possible
outcome of the system.  As $k_{mp}$ is increased further FP~3 undergoes
a subcritical Hopf bifurcation ($k_{mp} \approx 22$). Oscillations
around the unstable limit cycle surrounding FP~3 are interpreted as
{\em healthy} and are lost through a homoclinic bifurcation
($k_{mp} \approx 9$). Trajectories within the limit-cycle spiral into
FP~3 and are interpreted as {\em recurrent infection}.  Beyond
the Hopf bifurcation ($k_{mp} > 22$), FP~3 remains an unstable
spiral, up to very large $k_{mp}$ ($k_{mp} \approx  2000$) and
oscillations around FP 3 are interpreted as {\em healthy}. On
varying $k_{mp}$, only the position and eigenvalues of FP 3 are
changed. FP 5 remains unaffected which is interpreted as {\em
  persistent infectious 
inflammation} and remains a possible outcome of the system. If curve
(\ref{fp6}) is shallow enough ($b$ was small enough) so that FP 5 is
not yet created, then FP 3 remains the stable global attractor through
changes in $k_{mp}$.  (Inset :  In the $p$-$m$ plane, FP 3 varies in
position and  
stability as (\ref{fp6}) varies, causing the above
bifurcations. FP~5 remains unaffected.)}
\label{fig:varyer}
\end{figure}

Varying $k_{mp}$, the strength of the early response, does not affect the
stability or position of FP 5, the persistent non-infectious
inflammation point. Decreasing $k_{mp}$ takes a healthy state to a
state where healthy and recurrent infection co-exist and finally to
one of recurrent infection only as shown in
Fig.~\ref{fig:varyer}. Decreasing $k_{mp}$ below $1-(k_{lm}/k_l)f(0)$
makes FP 3  collide with FP 2 (the severely immuno-deficient state
where $p$ grows to saturation and $m$ and $l$ remain at background
values). In a transcritical bifurcation, FP 2 becomes the stable fixed
point of the system and FP 3 becomes unphysical since
$m<0$. Increasing $k_{mp}$ from the healthy value reduces the value of
$p$ in FP 3 - this reduces the range of initial conditions for healthy
behavior. However, for even very large values of $k_{mp}$, FP 3 is an
unstable spiral with complex eigenvalues supporting healthy behavior
for initial conditions starting close to FP 3.

Varying the strength of the late response, $b$ affects both FP 3 and
FP 5. As we decrease $b$, FP 3 undergoes a sub-critical Hopf followed
by the homoclinic bifurcation of the unstable limit cycle similar to
the above cases (Fig. \ref{fig:varylt}). FP 4 and FP 5 are created in
a saddle-node bifurcation when (\ref{fp6}) intersects $p=0$ by
increasing $b$. However, there is also an upper limit to $b$ beyond
which healthy behavior is not supported. The curve (\ref{fp6}) is
below zero when it intersects (\ref{fp2}) and FP 3 is unphysical. Thus
when the late response is too high, persistent non-infectious inflammation
is the only possible outcome.

In the preceding, although we have examined the behavior of the system by
varying one parameter at a time, it gives us a picture
of the global dynamics of the system. Given the strong nonlinear
saturation in the dynamics,  orbits can either approach a fixed point,
a limit cycle, or a strange attractor.  Given that FP~1 and FP~4 are
always unstable and FP 2 is stable only when $k_{mp}$ is very low,
this leaves FP 3 (or stable limit cycles around it) and FP 5 as the
only  candidates for global attractors. Ideally however, we would like
FP 3 to be an unstable spiral or be surrounded by an unstable limit
cycle, so that oscillations around it can be interpreted as healthy
behavior.

This gives an explanation for why healthy behavior always entails the
risk of uncontrolled inflammation.  In order for the pathogen to be
cleared, we require an unstable spiral that can take $p$ below
threshold.  However, if it were not for the (externally imposed)
elimination threshold the orbit would eventually end up at some
attractor and persistent non-infectious inflammation is the global
attractor. Conversely, if the curve (\ref{fp6}) has not yet
intersected $p=0$ to form FP 4 and FP 5 (no persistent inflammation
fixed point), FP 3 or a stable limit cycle around it (recurrent
infection) is the global attractor.

There is a possibility that FP 3 could undergo a supercritical Hopf
bifurcation (discussed in Appendix~\ref{appfp3}) which leads to a
stable limit cycle around it. However, this is generically interpreted
as recurrent infection because fine tuning would be required to ensure
that the limit cycle takes $p$ below threshold. Although, not possible
in our model, a strange attractor might occur in higher dimensional
models.  We believe that this is unlikely given the large amounts of
negative feedback and dissipation in the system.  However, even if it
were to exist, it may still not lead to a healthy situation as $p$ may
never be eliminated or the early and late inflammatory mediators could
stay elevated albeit in a chaotic manner.

Physiologically, this implies that in order to completely rid the body
of a pathogen, the inflammatory response must respond strongly and
remain elevated for a long enough time.  If it responds too weakly,
then the pathogen persists.  If it abates too quickly, then
there will be recurrent oscillatory infection.  However, if it
responds too strongly and too persistently then there is a risk that
it will be self-sustaining even after the pathogen is cleared. Thus,
there is a trade-off between being able to eliminate pathogens
completely and risking nonabating inflammation.   In
Appendix~\ref{appbzone}, we calculate the parameter ranges where a
healthy response is possible.

\section{Discussion}

Systemic inflammation and the ensuing organ damage
is a major cause of mortality today~\cite{tilney,eiseman}. This is also a disease
created by modern medicine.  Before the discovery of advanced
resuscitation techniques, patients
could not be kept alive long enough for the condition to fully
unfold. Patients often died from blood loss and severe infections
before uncontrolled inflammation arose. With the advent and
improvement of antibiotics and organ support therapy,
the condition has  become increasingly  relevant~\cite{bonerev96}. The
incidence of systemic  inflammation  is  also  expected to  increase
with  further advancement   of   medical   technology   and   the
aging of the population which are more susceptible~\cite{Angus}.

It  is  now  recognized  that  the  dysregulation  of  the  underlying
processes of  acute inflammation can lead to multiple organ  failure and
death and that  this is a common pathway  for diverse instigators such
as  trauma,  hemorrhage and  infection.   Our  simple  model of  acute
inflammatory response to infection shows the various negative outcomes
that  arise from  improper inflammatory  response. These  scenarios of
persistent     inflammation    (with    and     without    infection),
immuno-suppression and  recurrent infection have all  been observed in
critically  ill  patients~\cite{epidem02,bonerev96}.  Although  strict
correspondence with clinical reality is difficult to establish because
of the  simplicity of  our model,  we are able  to classify  our model
behavior  into  the same  broad  categories.   The  model has  a  rich
bifurcation  structure and exploring  it allows  us to  understand how
changing  parameters   can  take  the  system  from   one  outcome  to
another. The model  shows that in order to have  a healthy response to
infection the virulence of the pathogen cannot be too strong or too weak,
the early  pro-inflammatory response cannot  be too weak, and  the late
response cannot be too strong or  too weak.  Analysis of this model gives
clues  to  approach  the  problem  of treating  severe sepsis.  Vastly
different  therapeutic strategies  are  called for  to  deal with  the
diverse negative outcomes.

The strength  of the acute  inflammatory response varies  in different
individuals and may vary depending on age and environmental factors in
the same  individual.  Studies suggest  a strong genetic  influence on
the outcome of sepsis, and  genetics may explain the wide variation in
the individual response to infection~\cite{Genreview}. We examined the
effect of  the strength of early and  late pro-inflammatory responses.
The model suggests that only the strength of the late pro-inflammatory
wave  governs  predisposition  towards  a state  of  persistent
inflammation.  No  matter how  exuberant the early  wave may  be, only
controlling the feedback from the  late wave can determine whether the
outcome is  healthy recovery  or uncontrolled inflammation.   Thus any
therapy for persistent non-infectious inflammation must target the
slow pro-inflammatory mediators.  

Experiments have found that survival
was improved in infected mice when the moderately slow cytokine IL-6 was
reduced~\cite{riedemann}.  However, if too much IL-6 was removed then
there was a detrimental effect.  This result is consistent with our model in
that a small reduction lowered the possibility of a sustained
inflammatory response but lowering it too much precluded the
possibility of eliminating the bacteria.
Evidence also has suggested that down-regulating  HMGB-1, a  late 
acting  pro-inflammatory  mediator might  be  a  potential target  for
anti-sepsis therapies~\cite{HMGB}.  Activated Protein-C which has been
recently   approved  for  treatment,   is  also partially a  late
anti-inflammatory agent~\cite{APC}.    Previous therapeutic  attempts
have mostly   focused   on 
down-regulating the early  pro-inflammatory mediators and as predicted
by    the     model    have    not     shown    great    effectiveness
\cite{clintrial1,clintrial2}.  

On the other hand, if   the  patient  is
suffering  from persistent infectious  inflammation, then  therapies
must be  aimed at both  reducing  the  pathogen   load  and  the  late
pro-inflammatory response.  In this case, timing of the therapies may
be important.  It would be necessary to reduce the bacterial load
first before reducing the inflammation.

Conversely, low pathogen virulence or  a weak immune response can lead
to  low level  persistent or  recurrent infection.   Other theoretical
models  of infection  have  similar predictions.   Persistence of  the
tuberculosis  bacterium   {\it  Mycobacterium  tuberculosis}   at  low
densities for extended periods has  been suggested to be the result of
slow  growth rates~\cite{antia}.   Similarly, a  down-regulated immune
response to  {\it Helicobacter pylori} bacteria has  been suggested to
result    in    its    persistent    colonization   of    the    human
stomach~\cite{denise}.  In  the clinical setting, patients  in the ICU
with  decreased  host defenses  are  susceptible to  hospital-acquired
infections~\cite{nosoc}.   These   infections  which  may   be  easily
resolved  in   a  healthy   individual  might  result   in  unresolved
inflammation     and     prove      fatal     to     a     compromised
individual. Immuno-stimulatory  therapy might  be effective in  such a
situation.

The model suggests that a  healthy outcome is possible only when the
risk  of  persistent inflammation  is  also  present.   A strong  late
immune  response that  increases the  risk of  unabated inflammation
also ensures complete elimination of pathogen.  This could explain why
this risk has been retained by the evolution.

In the clinical setting, inflammatory states are defined by
symptoms and a few biological markers~\cite{sepsisdefnold}.  For example,
Systemic Inflammatory Response Syndrome (SIRS) is a condition
characterized by -
elevated respiratory and heart rates, fever, and an abnormal
white blood cell count. The severity of sepsis is based on the
presence of circulatory shock (low blood perfusion) and organ
failure~\cite{ASCC,muckart}.
Similarly, treatment and 
support of these critically ill patients is largely based on
clinical signs and a few biochemical and hematological parameters.
However, the
various outcomes of our model are  based on the levels of the immune
responders  and  the  pathogen.   Diverse  physiological  states  as
exemplified in  the model could  be manifested by the  same clinical
symptoms.   Hence,  one  major  problem for  finding  an  effective
treatment  of  sepsis is  a  diagnostic  one.   The model  shows  that
sepsis  should be  considered  as a  set  of distinct  physiological
disorders that require separate  therapies even though they may have
overlapping symptoms.  Categorizing septic states based on levels of
bio-markers rather than clinical symptoms would be the first step in
addressing   this  problem   and   effort  is   ongoing  in   that
direction~\cite{sepsisdefnnew,marshall}.

\section{Acknowledgments}
This work was funded in part by NIGMS grant R01-GM-67240 and the
A.~P.~Sloan Foundation.

\appendix
\section{Appendix}
\subsection{Eigenvalues of Fixed Points}
The Jacobian matrix of the model (\ref{model1}) - (\ref{model3}) is
\begin{equation}
\left(
\begin{array}{cccc}
k_p(1-2p)-k_{pm}m & -k_{pm}p & 0 \\ k_{mp}m(1-m) & (k_{mp}p+l)(1-2m)-1
& m(1-m)\\ 0 & \frac{k_{lm}}{w}{\rm sech}^{2}(\frac{m-\theta}{w}) &
-k_{l}
\end{array}
\right)
\label{jacobian}
\end{equation}

\subsubsection{FP 1}
\label{appfp1}
For FP1: $p=0,m=0,l=\frac{k_{lm}}{k_l}(1+\tanh(\theta/w))(\equiv
l^{0}$),  the Jacobian is 
\begin{equation}
\left(
\begin{array}{cccc}
k_p & 0 & 0\\ 0 & l^{0}-1& 0\\ 0 & \frac{k_{lm}}{w}{\rm
sech}^{2}\frac{\theta}{w}& -k_{l}
\end{array}
\right).
\end{equation}
This matrix is lower-triangular and the eigenvalues are  $k_p,
l^{0}-1, -k_{l}$. The fixed point is never stable as one of the
eigenvalues $k_{p}$ is always positive.

\subsubsection{FP 2}
\label{appfp2}
For FP2: $p=1,m=0,l=l^{0}$,  the Jacobian is
\begin{equation}
\left(
\begin{array}{cccc}
-k_p & -k_{pm} & 0\\ 0 & (k_{mp}+l^{0})-1& 0\\ 0 &
\frac{k_{lm}}{w}{\rm sech}^{2}\frac{\theta}{w}& -k_{l}
\end{array}
\right).
\end{equation}
The characteristic polynomial for the eigenvalue $\lambda$ is
\[(-k_p-\lambda)(-k_l-\lambda)(k_{mp}+l^0-1-\lambda)=0\]
with roots: $\lambda=-k_p$, $\lambda=-k_l$ and $\lambda=k_{mp}+l^0-1$.
Thus the condition for stability of FP2 is $k_{mp}<1-l^{0}$ or
$k_{mp}<1-\frac{k_{lm}}{k_l}(1-\tanh(\theta/w))$.  The other two
eigenvalues are always negative.

\subsubsection{FP 3}
\label{appfp3}
FP3 is the solution of the following equations,
\begin{eqnarray}
p&=&1-am\\ m&=&1-\frac{1}{k_{mp}p+l}\\ l&=&bf(m)
\end{eqnarray}
This may be reduced to a single transcendental equation for $m$:
\begin{equation}
m=1-\left(\frac{1}{k_{mp}}\left(\frac{1}{1-am}\right)+ bf(m)\right).
\end{equation}
All the parameter combinations  affect the position and eigenvalues of
FP 3.  Parameter choices for which $p$  or $m$ of FP 3 is  negative are
unphysical.  FP 3  could also have real or  complex eigenvalues.  FP 3
has  complex eigenvalues  and  can support  oscillatory behavior  when
(\ref{fp2}) intersects (\ref{fp6}) such that  $m$ of FP 3 is less than
$m$  of  FP  4  (Fig.~\ref{fig:nullclines}).  FP 3  can  undergo  Hopf
bifurcations in  this part of  the phase-space - the  subcritical Hopf
bifurcations on varying the  three parameter combinations are shown in
Figs. \ref{fig:varyp}, \ref{fig:varyer}, and \ref{fig:varylt}.
 
A supercritical Hopf bifurcation occurs when we change the coupling
curve (\ref{coupling}) so that $\theta$ is small (low threshold for
activation) and $w$ is large (shallow coupling curve).  In this case,
for increasing $b$ while keeping $a$ and $k_{mp}$ fixed at moderate
values, a state of persistent non-infectious inflammation will follow
recurrent infection and a small zone of stable limit cycles.  (Varying
the other parameters, for small $\theta$ and large $w$, does not
result in Hopf bifurcations). In the case of small $w$ (sharp coupling
curve) and small $\theta$ (low threshold of activation), it is
possible that more than one fixed point satisfies the conditions for
FP 3.  In this case, curve~(\ref{fp6}) intersects line (\ref{fp2})
twice before dipping below zero.  On varying the parameter
combinations, recurrent infection is bistable with persistent
inflammation and healthy behavior is never supported. These cases are
discussed in greater detail in Appendix~\ref{couplingcurve}.  In our
numerical examples, we have considered moderate $w$ and large
$\theta$, which result in subcritical Hopf bifurcations.

\subsubsection{FP 4 and FP 5}
\label{appfp45} 
These fixed points are the solutions of the following equations,
\begin{eqnarray}
p&=&0 \\ m&=&1-(\frac{1}{k_{mp}k_{pm}p+l})\\ l&=&\frac{k_{lm}}{k_l}f(m)
\end{eqnarray}
This may be reduced to one transcendental equation
\begin{equation}
bf(m)=\frac{1}{1-m}\label{fp45}
\end{equation}   
Equation (\ref{fp45}) may have 0, 1 or 2 solutions depending on the
parameters. Approximate solutions can be found for (\ref{fp45}) by
substituting a piece-wise linear function such as $f(m)=0,
m<\theta-w$; $f(m)=1+\frac{(m-\theta)}{w}, \theta-w<m<\theta+w$;
$f(m)=2, m>\theta+w$. Using the above approximate function and
assuming that FP 4 and 5 occur in region $\theta-w<m<\theta+w$, we
arrive at the following expressions for FP 4 and 5.
\begin{eqnarray}
p&=&0\\ m&=&\frac{q\mp\sqrt{q^2-\frac{4w}{b}}}{2}\\ l&=&bf(m)
\end{eqnarray}
where $q\equiv 1-\theta+w$.

This gives $b>4w/q^2$ as a lower  limit on $b$ for the existence of FP
4 and  5.  When  there are  two solutions, they  are formed  through a
saddle-node bifurcation as shown  in Fig.~\ref{fig:varylt}.  FP~4 is a
saddle point and FP~5 is a stable node.

The  eigenvalues of  FP  4 and  5  can be  found  by substituting  the
solutions     $m=m^0$    of     (\ref{fp45})    into     the    Jacobian
(\ref{jacobian}).       One      of      the       eigenvalues      is
$\lambda_1=k_{p}-k_{pm}m^0$  and the  other two  are functions  of the
parameters $k_l$,  $k_{lm}$, $\theta$ and  $w$. We note  that $k_{mp}$
does not appear in the  expressions for the position or eigenvalues of
FP 4 and 5.

\subsection{Regimes for Healthy Response}
\label{appbzone}
Trajectories can be interpreted as healthy when FP~3 supports unstable
oscillations as an unstable spiral or as a stable spiral surrounded by
an  unstable limit cycle.  The unstable  limit cycle  is lost  when it
undergoes a  non-generic homoclinic bifurcation when  it collides with
FP~1. The  regimes of the  various parameters (for fixed  $\theta$ and
$w$) which maintain healthy behavior are discussed below.

For healthy  behavior, $a$ should be  large enough so that  FP~3 is to
the     left    of     FP~5     in    Fig.~\ref{fig:nullclines}     or
$(1/a)<(q-\sqrt{q^2-4w/b})/2$. Increasing  $k_{mp}$ even over  a large
range, does not  make FP 3 unphysical, nor does  FP~3 lose its complex
eigenvalues - thus even  for very large $k_{mp}$, healthy oscillations
are  supported.  The  upper  limit for  $a$  and the  lower limit  for
$k_{mp}$  to maintain healthy  oscillations are  difficult to  find in
closed  form as these  variations result  in the  radius of  the limit
cycle  increasing and  oscillations being  lost through  a non-generic
homoclinic bifurcation.

The strength of the late immune response $b$ should not be too high or
too low.  If $b$ is too  high, then FP  3 has $p<0$ and  hence healthy
behavior cannot be supported. In order for the late immune response to
not  be too  high, we  require that  $p$ of  condition  (\ref{fp6}) be
positive so that FP 3  is not unphysical. Applying this to (\ref{fp6})
gives the condition
\begin{equation}
b<\frac{1}{1-\tanh(\theta/w)}.
\label{w1}
\end{equation}
If $b$ is  too low, FP 4 and  5 are not yet formed  and (\ref{fp6}) is
well above $p=0$ so FP 3  remains the global attractor and is a stable
spiral. To avoid  too low of a response,  we would like FP 4  and 5 to
exist so that unstable oscillations from FP 3 are eventually attracted
to  FP 5.   Thus  the  condition for  existence  for FP  4  and 5,  as
calculated above in \ref{appfp45}, is
\begin{equation}
b>\frac{4w}{(1-\theta+w)^2}.
\label{w2}
\end{equation}

\subsubsection{Effect of Varying the Coupling Curve}
\label{couplingcurve}
The  shape  of the  coupling  curve (\ref{coupling})  (i.e.~parameters
$\theta$  and $w$)  can alter  the  results.  We  used $\theta=1$  and
$w=0.5$ in  our analysis. Parameter $\theta$ sets  the threshold where
$l$  is activated  by $m$  and parameter  $w$ gives  the  steepness of
recruitment.

From numerical simulations,  we find that when $w$  is too small, FP~3
never undergoes a Hopf  bifurcation, and as $k_{lm}/k_l$ is increased,
recurrent  infection and  persistent inflammation  are  bistable (For
example, $w=0.005$ 
with 'healthy' parameter set in Fig.~\ref{fig:healthy}).  As discussed
in Appendix~\ref{appfp3},  small $w$ could result in a shallow enough
curve (\ref{fp6}) that intersects line  (\ref{fp2})  more than once.
The two fixed points replacing FP~3
arise in a saddle-node  bifurcation. The  stable  node is
bistable  with FP~5  implying that recurrent  infection  is bistable  with
persistent non-infectious inflammation. 
When $w$ is too large ($w=1$),
FP~3 is a stable spiral that becomes unphysical as $k_{lm}/k_l$ increases.
Thus the system supports recurrent infection followed by
persistent inflammation and healthy behavior does not exist.
In summary, for $w$ too small or large the subcritical Hopf
bifurcation of FP 3 (and hence healthy behavior) 
does not occur for any of the parameter combinations.

When $\theta$  is small ($\theta=0.1$), FP 3  undergoes a supercritical
Hopf  bifurcation as  $b$ is  varied  but this  cannot be  generically
interpreted as  healthy behavior as oscillations about  a stable limit
cycle  may  not always  take  $p$  below  threshold.  As  $\theta$  is
increased, subcritical  Hopf bifurcations  are possible. The  range of
parameter $b$,  given by (\ref{w1}) and  (\ref{w2}), where oscillatory
behavior is  observed also increases as $\theta$  is increased.  Thus,
large $\theta$ and moderate $w$  would maximize the region for healthy
response.

In  the   above  analysis,  we   have  looked  at   varying  parameter
combinations rather than individual  parameters for conciseness of the
discussion. Varying the parameters  such as $k_p$ and  $k_{pm}$ individually
does  not  change  the  bifurcation  diagrams  qualitatively  and  our
interpretations   of   the  effect   of   pathogen  susceptibility   remain
unchanged. Similarly, varying $k_{lm}$ and $k_l$ individually does not
change  our  conclusions about  the  strength  of  late response.  The
phase-space  analyzed is  restricted between  $0<m<1$  since condition
(\ref{fp6}) prevents  $m$ from increasing beyond one.  Using any other
function  such  as $1/(1+m)$  to  saturate  $m$  can change the upper
bound of $m$ but the same behavior is retained.

\bibliographystyle{jtb}  \bibliography{jtb}

\newpage
\begin{figure} 
\begin{center}
\includegraphics[angle=270,scale=0.4]{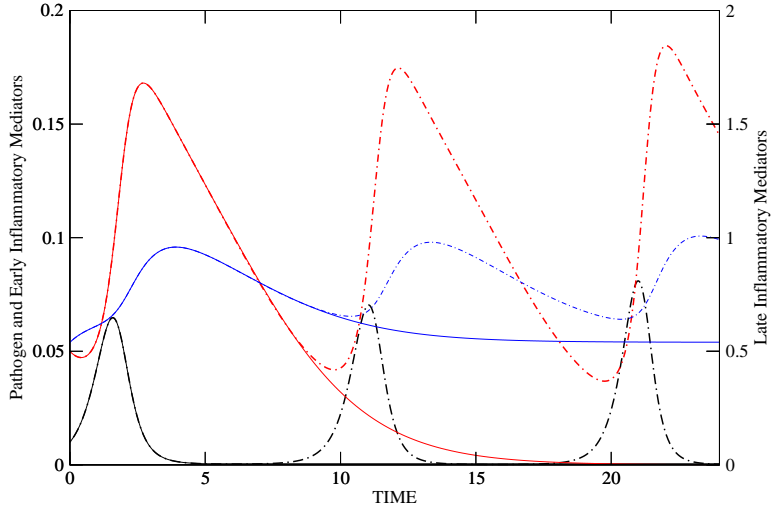}
\end{center}
\caption{Time courses of $p$ (black), $m$ (red) and $l$ (blue) for the {\it healthy} response.  We
show orbits  with the $p$ elimination threshold (solid line) and
without (dotted line). With the threshold (set at $p_0=0.0005$), when $p$ drops below $p_0$, $p$ is set to zero and $m$ and $l$ return to background values.  Without the threshold, the orbits spiral outwards.
Parameters used
are ${k_{pm}}=30, {k_p}=3, k_{mp}=25, {k_{lm}}=15, {k_l}=1$. Initial
conditions are $p(0)=0.01, m(0)=0.05, l(0)=0.539$.}
\label{fig:healthy}
\end{figure}

\end{document}